\RequirePackage{lineno}
\documentclass[prd,twocolumn,nofootinbib,superscriptaddress,floatfix]{revtex4}
\usepackage{cancel}
\usepackage{graphicx}
\usepackage{epsfig}
\usepackage{amsmath,amsfonts,amssymb}
\usepackage{t1enc}

\begin{document}


\title{Flavoured searches for type-III seesaw at the LHC}

\author{J.A. Aguilar-Saavedra}
\affiliation{Departamento de F\'{\i}sica Te\'orica y del Cosmos, Universidad de Granada,
 Granada, Spain}
 \affiliation{Departamento de Fisica, Universidade de Coimbra,  Coimbra, Portugal}
\author{P.M. Boavida}
\author{F.R. Joaquim}
\affiliation{Departamento de F\'{\i}sica and CFTP, Instituto Superior T\'ecnico, Universidade de Lisboa, Lisboa, Portugal}

\begin{abstract}
We present a reinterpretation of CMS searches for type-III seesaw lepton triplets at the LHC, both for the normal and the inverse seesaw. We find that, in contrast with previous expectations, these searches in the trilepton final state have a good sensitivity to triplets with predominant coupling to the $\tau$ lepton. We also show that the limits resulting from direct searches can be neatly presented for arbitrary masses and general flavour mixings. Thus, it turns out that the common (and often unrealistic) simplifying assumptions about the flavour couplings of the triplets used by experimental collaborations to present their results are unnecessary and should be dropped.
\end{abstract}

\maketitle

At present, the only solid manifestation of physics beyond the standard model (SM) are neutrino oscillations, which require neutrinos to be massive. Among the proposed theoretical frameworks that could account for neutrino masses, those relying on neutrino mass suppression via the decoupling of new heavy degrees do freedom (seesaw mechanism) have become the simplest paradigm for neutrino mass generation. If light enough, the {\em seesaw mediators} could possibly leave signatures at high-energy colliders like the Large Hadron Collider (LHC). Heavy neutrino singlets, such as those introduced in type-I seesaw~\cite{Minkowski:1977sc,GellMann:1980vs,Yanagida,Glashow,Mohapatra:1979ia}, can be produced in association with a charged lepton, but the small mixing with the SM particles~\cite{Nardi:1994iv,Tommasini:1995ii,Kersten:2007vk,delAguila:2008pw,delAguila:2008ks} leads to hardly observable production rates for the new fermions~\cite{delAguila:2007em} (pair production is even more suppressed except in the presence of new neutral bosons~\cite{delAguila:2007ua,Huitu:2008gf,AguilarSaavedra:2009ik,Perez:2009mu}). On the other hand, scalar triplets of a type-II seesaw~\cite{Magg:1980ut,Barbieri:1979ag,Cheng:1980qt,Schechter:1980gr,Mohapatra:1980yp,Lazarides:1980nt} and triplet fermions of a type-III seesaw~\cite{Foot:1988aq} can be produced in pairs via electroweak gauge interactions, with cross sections that could allow for their observation at the LHC~\cite{Huitu:1996su,Akeroyd:2005gt,Akeroyd:2007zv,Franceschini:2008pz,Perez:2008zc,delAguila:2008cj,Arhrib:2009mz,Akeroyd:2010ip}.

Collider signals of heavy lepton triplets  (or heavy neutrinos of a type-I seesaw) that mainly couple to the $\tau$ lepton are difficult to detect, as they produce one or more $\tau$'s as decay products (see for example~\cite{delAguila:2008cj}). The subsequent decay products of these $\tau$ leptons that can be seen in the detectors are either electrons and muons (with a branching ratio around $1/3$) or narrow jets. In the former case, these secondary leptons from $\tau$ decay are less energetic than the parent $\tau$, and the resulting signals are usually difficult to extract from SM backgrounds. In the latter, the signals suffer from even larger backgrounds from the production of top pairs, weak bosons plus jets, etc. This fact partially explains why most experimental searches by the CMS and ATLAS Collaborations assume that heavy neutrinos only couple to the electron and/or muon. Specifically,
expressing the charged-current interactions between the light leptons $\ell_i=e,\mu,\tau$ and heavy neutrinos $N_j$, $j=1,2,3$ (with $N\equiv N_1$ the lightest one) as
\begin{equation}
\mathcal{L} = -\frac{g}{\sqrt 2} V_{\ell_i N_j} \bar \ell_{iL} \gamma^\mu N_{jL} W_{\mu}^{-} + \text{H.c.} \,,
\end{equation}
the CMS Collaboration considers~\cite{CMS:2012ra} three benchmark scenarios with (i) $V_{eN} \neq 0$, $V_{\mu N} = V_{\tau N} = 0$; (ii) $V_{\mu N} \neq 0$, $V_{e N} = V_{\tau N} = 0$; (iii)
$V_{eN} = V_{\mu N} = V_{\tau N}$. A mixing $V_{\tau N}=0$, although not excluded, is neither natural nor has any theoretical motivation, given the observed mixing in the light neutrino sector. Indeed, in the basis where the light charged lepton mass matrix is diagonal, the neutrino mass matrix $m_\nu$ and charged interaction mixings resulting from the seesaw mechanism are
\begin{align}
& (m_\nu)_{ij} = - \frac{v^2}{2} \frac{Y_{ik} Y_{jk}}{m_{N_k}} \,,
&& V_{\ell_i N_j} = \frac{v}{\sqrt 2} \frac{Y_{ij}}{m_{N_j}} \,, 
\label{ec:seesaw}
\end{align}
where $i,j,k=1,2,3$, and a sum over $k$ is implied in the first equation; $v=246$~GeV is the Higgs vacuum expectation value, $m_{N_i}$ are the heavy neutrino masses and $Y_{ij}$ is a $3 \times 3$ matrix of Yukawa couplings, whose precise definition (different in type-I and type-III) is not important here. In this basis, the neutrino mass matrix $m_\nu$ exhibits large off-diagonal entries, suggesting that $Y_{ij}$ is not diagonal but, in general, that heavy neutrinos do mix with $e,\mu,\tau$. A flavour benchmark $V_{eN} = V_{\mu N} = V_{\tau N}$ is more realistic than those with $V_{\tau N}=0$ but it only represents one particular point in a two-dimensional parameter space.

\begin{figure}[htb]
\begin{center}
\epsfig{file=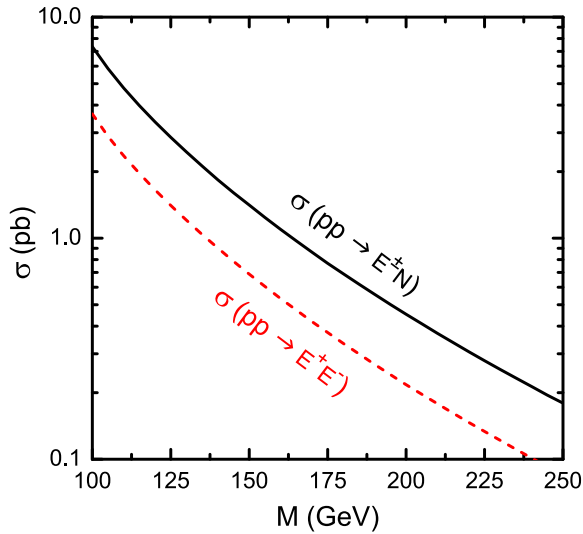,height=7cm,clip=} 
\caption{Cross section for pair production of heavy leptons in type-III seesaw, for a CM energy of 7 TeV.}
\label{fig:xsec}
\end{center}
\end{figure}

Experimentally, there are few cases in which the sensitivity to final states involving $\tau$ leptons is good. One of these was pointed out in previous works~\cite{AguilarSaavedra:2012fu,AguilarSaavedra:2012gf}. The ATLAS and CMS Collaborations have searched for
heavy neutrino production in type-I seesaw mediated by a heavy $W_R$ boson, in the context of left-right models~\cite{ATLAS:2012ak,Chatrchyan:2012fla}, assuming that heavy neutrinos couple only to the electron or muon. However, for $W_R$ boson and heavy neutrino masses probed by the experiments, $M_{W_R} \lesssim 2.5$ TeV, $m_N \lesssim 1.5$ TeV, the produced charged leptons are very energetic. Thus, even if these are $\tau$'s, the daughter leptons from their decay are still energetic enough to easily pass the experimental selections in~\cite{ATLAS:2012ak,Chatrchyan:2012fla} with a good efficiency. The only penalty for the $\tau$ signals in this case is the $\tau$ leptonic branching ratio. Then, the experimental searches can be straightforwardly extended to a general mixing as it has been done in~\cite{AguilarSaavedra:2012fu}.
A second interesting and complementary example is presented here, and concerns the production of heavy neutral ($N$) and charged ($E$) leptons in a type-III seesaw, which can be pair-produced in $pp$ collisions. The corresponding processes and decay modes are
\begin{align}
& pp \to E^\pm N \,, && E^\pm  \to \nu_i W^\pm / \ell_i^\pm Z / \ell_i^\pm H \,, \notag \\
&&& N \to \ell_j^\pm W^\mp / \nu_j Z / \nu_j H \,, \notag \\
& pp \to E^+ E^- \,, && E^+  \to \nu_i W^+ / \ell_i^+ Z / \ell_i^+ H \,, \notag \\
&&& E^-  \to \nu_j W^- / \ell_j^- Z / \ell_j^- H \,,
\label{ec:pairM}
\end{align}
with $i,j=1,2,3$, $(\ell_1,\ell_2,\ell_3)=(e,\mu,\tau)$. The total production cross sections only depend on the masses, which are nearly equal, $m_N \simeq m_E \equiv M$, and are presented in Fig.~\ref{fig:xsec}, for a centre of mass (CM) energy of 7 TeV.
For triplet masses probed at the LHC, $M \sim 200$ GeV, the charged leptons $\ell_i ,\ell_j$ in the final state are not so energetic. But the trilepton final state in which these signals have been searched for in~\cite{CMS:2012ra} is very clean~\cite{delAguila:2009bb}, and seesaw signals could be spotted over the SM backgrounds even if they produce leptons that are not very energetic. Three leptons can be produced in various decay chains of $E^\pm N$ and $E^+ E^-$ pairs, but the most important one, with a branching ratio around 7\%, is
\begin{equation}
E^\pm \to \ell_i^\pm Z/ \ell_i^\pm H \quad N \to  \ell_j^\pm W^\mp \,,
\end{equation}
with leptonic decay of the $W$ boson and hadronic (or invisible) decay of $Z$ and $H$. Other modes, including those where two opposite-charge leptons result from $Z$ decay, have much smaller branching ratios. Note that in the trilepton final state one of the three leptons necessarily results from the decay of a $W$ or $Z$ boson and its flavour is not related to the lepton triplet couplings; among the two others, $\ell_i,\ell_j$, one is often an electron or muon unless the triplet couplings to these leptons are very suppressed. Therefore, unless $V_{\tau N} \gg  V_{eN},V_{\mu N}$, two of the leptons are energetic enough to allow the event to pass the analysis selection. In this work we take advantage of this fact to provide a flavoured reinterpretation of the CMS search for type-III seesaw in trilepton final states~\cite{CMS:2012ra}, considering arbitrary flavour mixing of the heavy leptons. Moreover, we also consider an inverse type-III seesaw~\cite{delAguila:2008hw}. This is a variant in which, as it happens in inverse type-I seesaw~\cite{Mohapatra:1986bd} the heavy neutrino $N$ is a Dirac particle. But, at variance with it, there exist two heavy charged leptons $E_1^-$, $E_2^+$. In this scenario the heavy leptons have the same production processes as in the Majorana case but the decays are different. The production processes and allowed decays are
\begin{align}
& pp \to E_1^+ N \,, && E_1^+  \to \ell_i^+ Z / \ell_i^+ H \,, \notag \\
&&& N \to \ell_j^- W^+ / \nu_j Z / \nu_j H \,, \notag \\
& pp \to E_2^+ \bar N \,, && E_2^+  \to \nu_i W^+  \,, \notag \\
&&& \bar N \to \ell_j^+ W^- / \nu_j Z / \nu_j H \,, \notag \\
& pp \to E_1^+ E_1^- \,, && E_1^+  \to \ell_i^+ Z / \ell_i^+ H \,, \notag \\
&&& E_1^-  \to \ell_j^- Z / \ell_j^- H \,, \notag \\
& pp \to E_2^+ E_2^- \,, && E_2^+  \to \nu_i W^+  \,, \notag \\
&&& E_2^-  \to \nu_j W^- \,,
\label{ec:pairD}
\end{align}
together with the charge conjugate of the first two. (Notice that $E_2^+ E_2^-$ does not yield three leptons in the final state.) The cross sections are the same as in the minimal type-III seesaw shown in Fig.~\ref{fig:xsec}, $\sigma(E_1^\pm N) = \sigma(E_2^\pm N) = \sigma (E^\pm N)$; $\sigma(E_1^+ E_1^-) =  \sigma(E_2^+ E_2^-) = \sigma(E^+ E^-)$. Hence, the size of the trilepton signal is approximately twice larger than for a Majorana triplet. 

In order to estimate the sensitivity of the existing type-III seesaw search to general flavour mixings, beyond the benchmarks used by the CMS Collaboration, we have performed a fast simulation analysis of the processes in Eqs.~(\ref{ec:pairM}) and (\ref{ec:pairD}), including all possible decays,
 using {\sc Triada}~\cite{AguilarSaavedra:2009ik} for the signal generation, with four benchmark values of the triplet mass, $M = 100,150,200,250$ GeV. For each mass we have simulated nine samples per process, corresponding to $i,j=1,2,3$. We have used {\sc Pythia}~\cite{Sjostrand:2006za} for hadronisation and {\sc PGS 4}~\cite{pgs4} for the simulation of a generic LHC detector. To mimick the experimental analysis, we have applied the selection cuts in~\cite{CMS:2012ra}. For simplicity, we do not split trilepton events into classes by flavour ($e^+ e^+ e^-$, $e^+ e^+ \mu^-$, etc.) but sum all events that pass the analysis selection.\footnote{Notice that the CMS analysis only selects trilepton events with a total charge of $+1$, hereby reducing the signal to 60\% of the trilepton events resulting from the processes in Eqs.~(\ref{ec:pairM}).} By comparing the numbers of expected signal events for $M=200$ GeV, $i,j=1$ (lepton triplet coupling only to $e$), and $M=200$ GeV, $i,j=2$ (coupling only to $\mu$) between our fast simulation and the more realistic one in~\cite{CMS:2012ra} we find that the additional efficiencies ($e$ and $\mu$ detection efficiencies, trigger, etc.) that our fast simulation does not implement can be taken into account with an additional efficiency factor $\epsilon_0 \simeq 0.5$ that is nearly the same for both channels.

For each production process in Eqs.~(\ref{ec:pairM}), (\ref{ec:pairD}) and each choice of $i,j$, the efficiency $\epsilon_{ij}$ can be defined as $\epsilon_{ij}= N_{ij} / N^0_{ij}$, with $N^0_{ij}$ being the number of simulated events and $N_{ij}$ the number of events that pass the selection cuts (including the global efficiency $\epsilon_0$). These are collected in Tables~\ref{tab:eff1} and~\ref{tab:eff2} for the Majorana and Dirac cases, respectively. The numbers are relative to the total production cross section. Notice that the $EN$ efficiencies are not symmetric, $\epsilon_{ij} \neq \epsilon_{ji}$, because the first charged leptons results from the decay of $E$ and the second one from the decay of $N$. For $E^+ E^-$ production the small differences $\epsilon_{ij} \neq \epsilon_{ji}$ are due to the Monte Carlo statistics. (The uncertainty in our results due to the statistics of our Monte Carlo simulations has been ignored, as it can be eventually reduced with larger samples.) 
\begin{table}[htb]
\caption{Efficiencies (in percent) for the trilepton selection in the case of a Majorana lepton triplet. Values smaller than $10^{-2}$ are indicated by a dash.
\label{tab:eff1}}
\begin{center}
\begin{tabular}{ccccccccccccc}
\hline
\hline
$M$& \multicolumn{2}{c}{$100$ GeV} & \multicolumn{2}{c}{$150$ GeV} & \multicolumn{2}{c}{$200$ GeV} & \multicolumn{2}{c}{$250$ GeV} \\
$\ell_i \ell_j$ & $E^\pm N$ & $E^+ E^-$  & $E^\pm N$ & $E^+ E^-$ & $E^\pm N$ & $E^+ E^-$ & $E^\pm N$ & $E^+ E^-$ & \\
$	ee	 $	&	0.19	&	 --	    &	0.36	&	0.06	&	0.46	&	0.10	&	0.52	&	0.12	\\
$e \mu	 $	&	0.16	&	 --  	&	0.35	&	0.06	&	0.45	&	0.10	&	0.43	&	0.11	\\
$e \tau	 $	&	0.03	&	0.01	&	0.11	&	0.05	&	0.14	&	0.07	&	0.18	&	0.09	\\
$\mu e	 $	&	0.21	&	0.01	&	0.39	&	0.06	&	0.46	&	0.10	&	0.48	&	0.11	\\
$\mu \mu $	&	0.19	&	0.01	&	0.36	&	0.06	&	0.42	&	0.11	&	0.42	&	0.10	\\
$\mu \tau$	&	0.03	&	0.01	&	0.12	&	0.05	&	0.15	&	0.08	&	0.17	&	0.08	\\
$\tau e  $	&	0.16	&	 --  	&	0.20	&	0.03	&	0.24	&	0.04	&	0.28	&	0.06	\\
$\tau \mu$	&	0.13	&	 -- 	&	0.19	&	0.03	&	0.19	&	0.05	&	0.24	&	0.06	\\
$\tau\tau$	&	0.02	&	 -- 	&	0.06	&	0.01	&	0.08	&	0.02	&	0.09	&	0.02	\\
\hline
\hline
\end{tabular}
\end{center}
\end{table}
\begin{table*}[htb]
\caption{The same as Table~\ref{tab:eff1}, for the relevant processes in the case of a heavy Dirac lepton triplet.
\label{tab:eff2}}
\begin{center}
\begin{tabular}{ccccccccccccc}
\hline
\hline
$M$& \multicolumn{3}{c}{$100$ GeV} &  \multicolumn{3}{c}{$150$ GeV} & \multicolumn{3}{c}{$200$ GeV} & \multicolumn{3}{c}{$250$ GeV} \\
$\ell_i \ell_j$ & $E_1^\pm N$ & $E_2^\pm N$ & $E_1^+ E_1^-$ 
                    & $E_1^\pm N$ & $E_2^\pm N$ & $E_1^+ E_1^-$
                    & $E_1^\pm N$ & $E_2^\pm N$ & $E_1^+ E_1^-$
                    & $E_1^\pm N$ & $E_2^\pm N$ & $E_1^+ E_1^-$ \\
$	ee	$	&	0.35	&	0.15	&	0.04	&	0.75	&	0.19	&	0.18	&	0.80	&	0.24	&	0.27	&	0.79	&	0.27	&	0.26	\\
$e\mu	$	&	0.33	&	0.14	&	0.04	&	0.84	&	0.16	&	0.22	&	0.80	&	0.19	&	0.27	&	0.71	&	0.20	&	0.28	\\
$e\tau	$	&	0.04	&	0.02	&	0.03	&	0.25	&	0.04	&	0.12	&	0.28	&	0.05	&	0.15	&	0.29	&	0.08	&	0.15	\\
$\mu e	$	&	0.37	&	0.16	&	0.05	&	0.83	&	0.19	&	0.21	&	0.77	&	0.24	&	0.26	&	0.73	&	0.26	&	0.24	\\
$\mu\mu	$	&	0.34	&	0.14	&	0.04	&	0.71	&	0.17	&	0.18	&	0.73	&	0.20	&	0.27	&	0.69	&	0.21	&	0.25	\\
$\mu\tau$	&	0.06	&	0.02	&	0.05	&	0.25	&	0.05	&	0.12	&	0.28	&	0.06	&	0.16	&	0.28	&	0.07	&	0.16	\\
$\tau e	$	&	0.07	&	0.16	&	0.01	&	0.23	&	0.18	&	0.05	&	0.27	&	0.22	&	0.09	&	0.27	&	0.27	&	0.10	\\
$\tau\mu$	&	0.07	&	0.14	&	0.01	&	0.20	&	0.18	&	0.05	&	0.27	&	0.19	&	0.09	&	0.27	&	0.20	&	0.10	\\
$\tau\tau$	&	0.02	&	0.02	&	0.01	&	0.07	&	0.04	&	0.03	&	0.09	&	0.07	&	0.04	&	0.09	&	0.07	&	0.05	\\

\hline
\hline
\end{tabular}
\end{center}
\end{table*}
For masses other than $M=100,150,200,250$ GeV the efficiencies are obtained by a linear interpolation (or extrapolation) between the simulated points.
Then, the number of signal events for a given process and arbitrary mixings $V_{eN}$, $V_{\mu N}$, $V_{\tau N}$ can be calculated as
\begin{equation}
N = L \sigma \sum_{ij} \epsilon_{ij} v_{\ell_i N}^2 v_{\ell_j N}^2 \,,
\end{equation}
with $\sigma$ the total cross section, $L=4.9~\text{fb}^{-1}$ the luminosity and
\begin{equation}
\quad v_{\ell_i N} = \frac{|V_{\ell_i N}|}{\sqrt{|V_{eN}|^2 + |V_{\mu N}|^2 + |V_{\tau N}|^2}}  \,.
\end{equation}
The total number of signal events is obtained by summing over the contributions of the different processes in Eqs.~(\ref{ec:pairM}) for the Majorana case and Eqs.(\ref{ec:pairD}) for the Dirac case. The 95\% confidence level exclusion limits are computed by a simple $\chi^2$ for the number of expected events, summing the statistical and systematic uncertainties in quadrature. We also sum the systematic uncertainties of the different trilepton channels in quadrature, which amounts to neglecting the correlations between them. These systematic uncertainties arise from the same sources: normalisation of the various backgrounds, electron and muon efficiencies, etc.\ that contribute with different weights to the total uncertainty in the different channels. Then, one expects some small positive correlations, which are not available from the experimental analysis. Neglecting them thus leads to slightly optimistic limits.\footnote{For example, the CMS search excludes $M \leq 200$ GeV for a triplet coupling only to the electron, while the limit obtained from our simulation is slightly higher, $M \leq 204$ GeV.}

\begin{figure}[htb]
\begin{center}
\begin{tabular}{c}
\epsfig{file=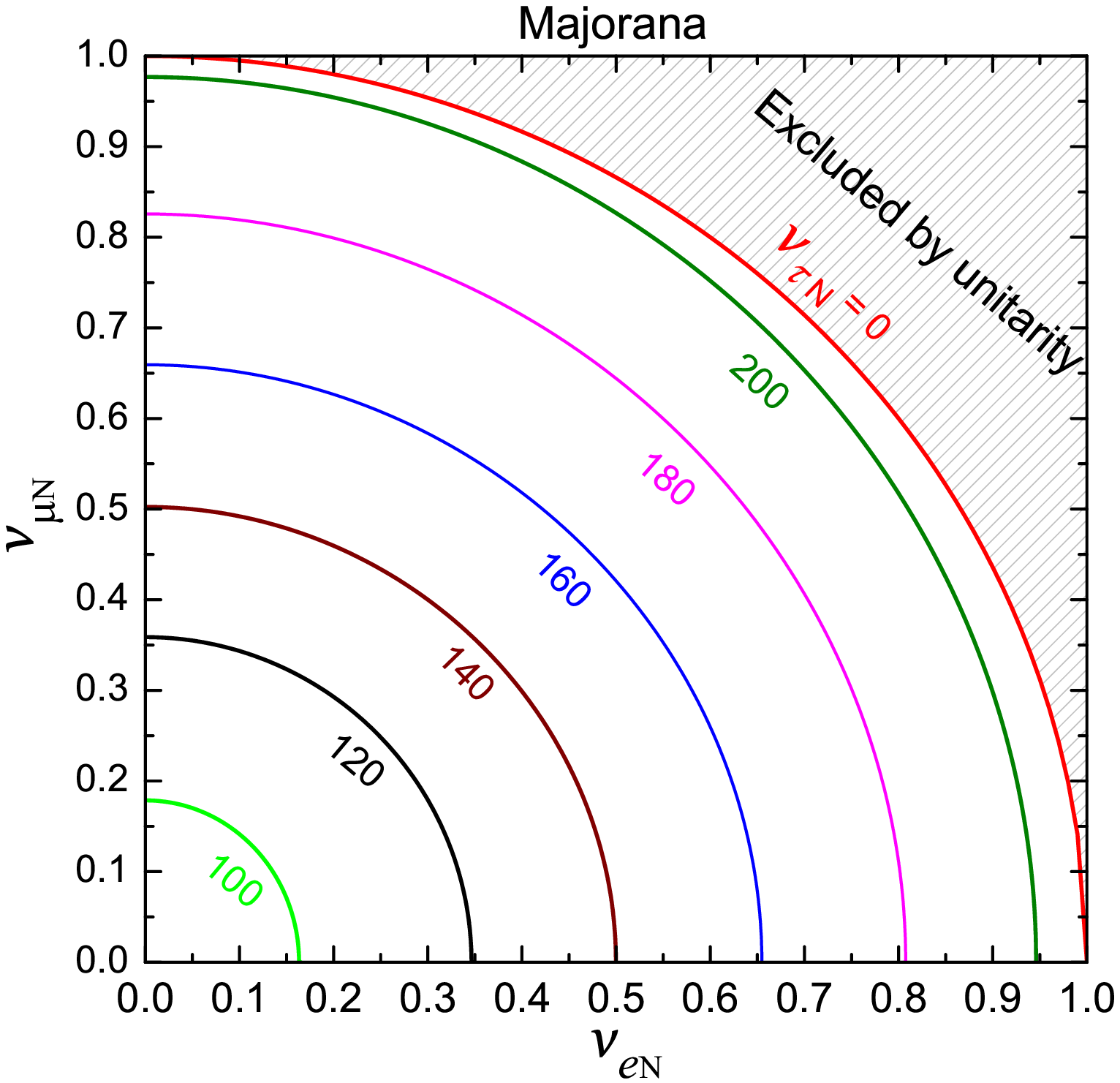,height=7cm,clip=} \\[2mm]
\epsfig{file=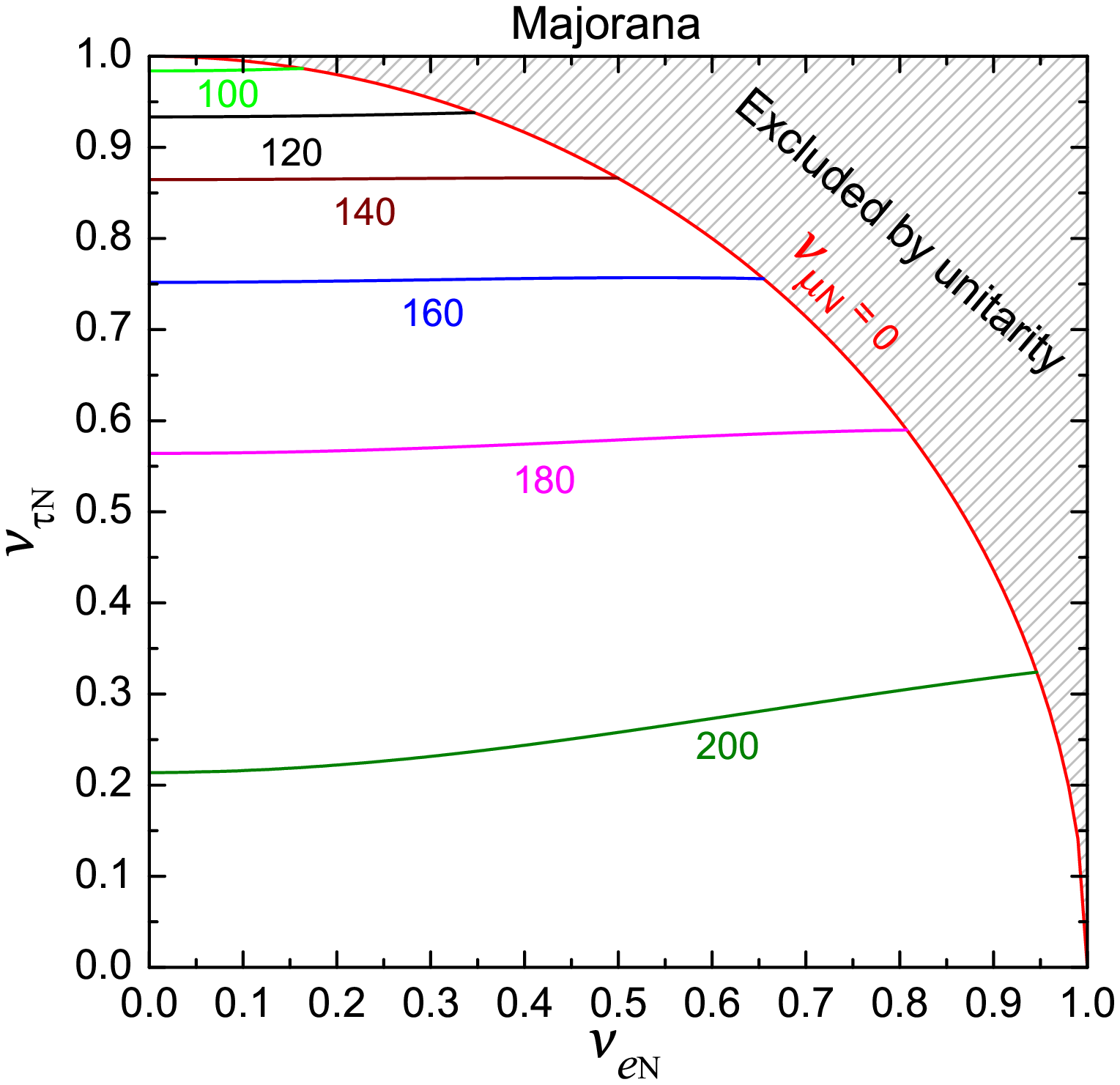,height=7cm,clip=} 
\end{tabular}
\caption{General limits on heavy Majorana triplets, as a function of $v_{eN}, v_{\mu N}$ and the triplet mass (top), or as a function of $v_{eN}, v_{\tau N}$ and the triplet mass (bottom). In the upper plot the excluded regions for a given $M$ range from the corresponding contour to the outer arc, while in the lower plot the excluded regions lie from the corresponding contour to the lower axis $v_{\tau N}=0$. The masses in the labels are in GeV. }
\label{fig:lim1}
\end{center}
\end{figure}

\begin{figure}[htb]
\begin{center}
\begin{tabular}{c}
\epsfig{file=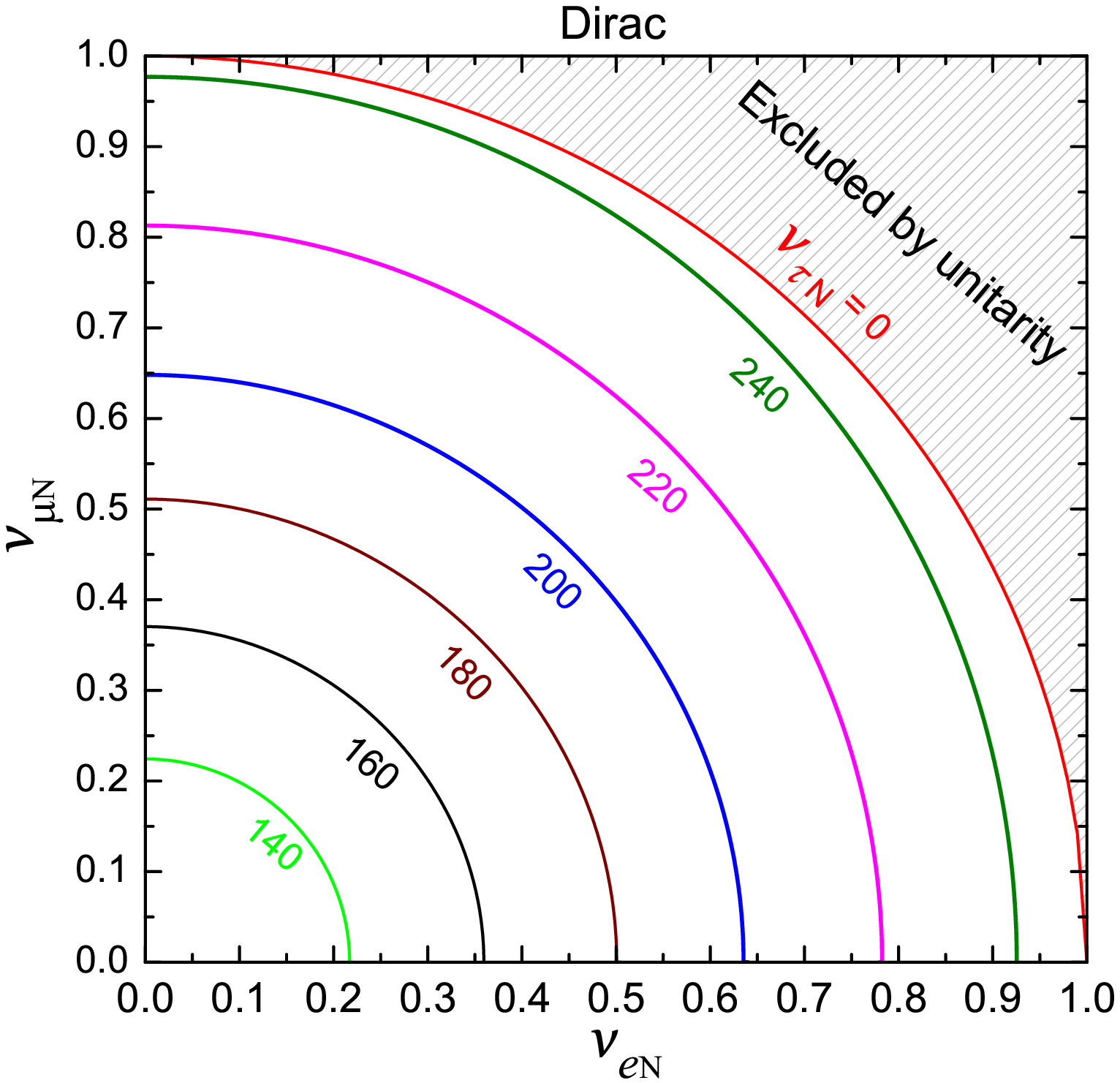,height=7cm,clip=} \\[2mm]
\epsfig{file=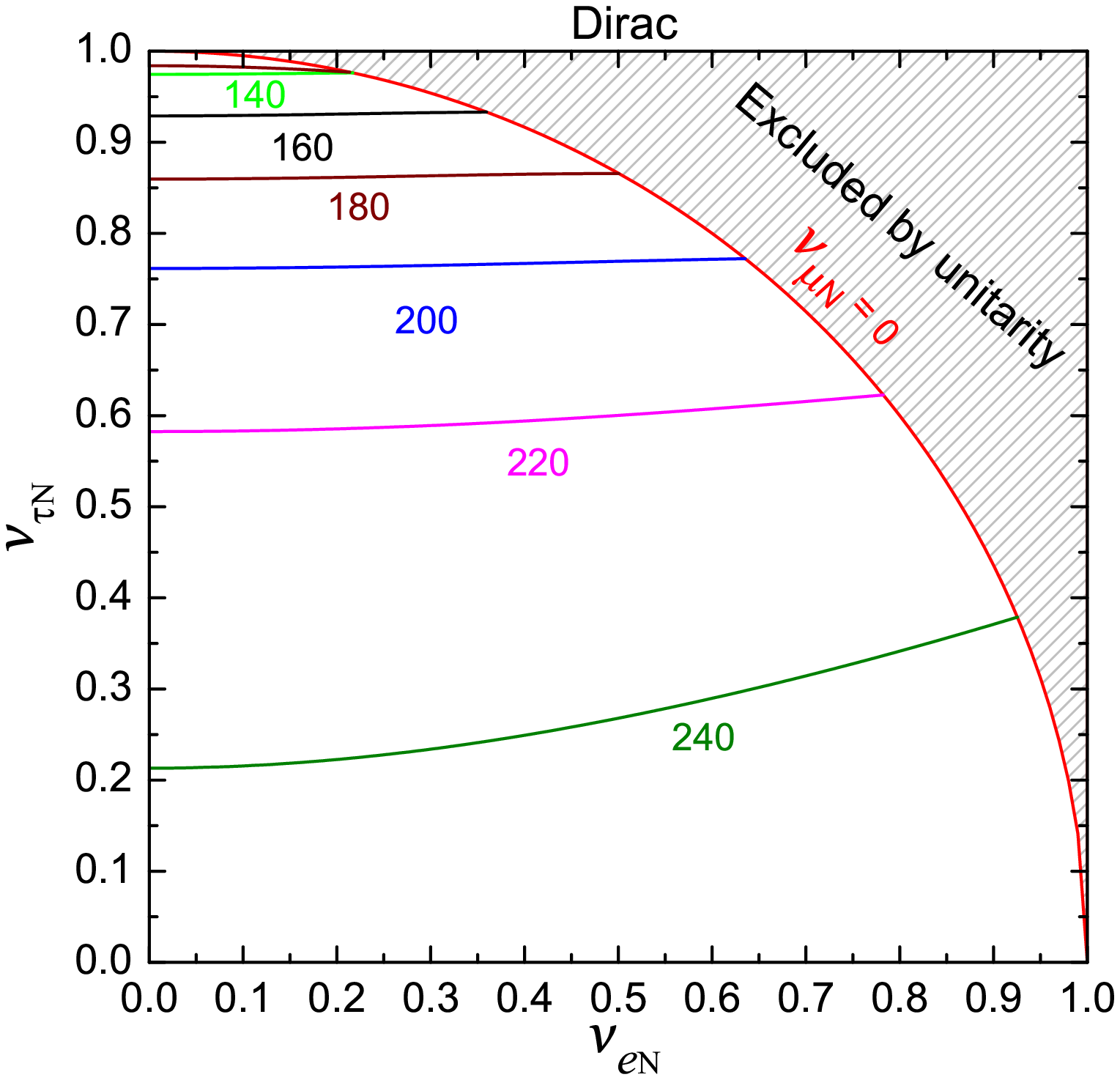,height=7cm,clip=} 
\end{tabular}
\caption{The same as Fig.~\ref{fig:lim1}, for a Dirac triplet. }
\label{fig:lim2}
\end{center}
\end{figure}

The results of our analysis can be neatly presented in terms of two-dimensional exclusion plots in flavour space, and simultaneously for different triplet masses. Since by definition we have a unitarity condition
\begin{equation}
 v_{eN}^2 + v_{\mu N}^2 + v_{\tau N}^2 = 1 \,,
\end{equation} 
one can choose any two of these variables to parameterise the heavy triplet mixing, and the third one is obtained using the above equation.
Figure~\ref{fig:lim1} shows the excluded parameter space for a Majorana triplet, using either $v_{eN}, v_{\mu N}$ (top) or $v_{eN}, v_{\tau N}$ (bottom) as independent variables, and showing the 95\% exclusion limits for different triplet masses between 100 and 250 GeV. (Note that the plots on the top and bottom panels contain the same information but shown from a different perspective.) Figure~\ref{fig:lim2} does the same for a Dirac triplet.
In both cases we observe that the flavour dependence of the limits is mainly given by the value of $v_{\tau N}$, that is, the lines in the two upper plots (Majorana and Dirac) are almost circular arcs and the lines in the lower plots are almost flat. The limits corresponding to different masses do not overlap, since the exponential decrease of the cross section with mass (see Fig.~\ref{fig:xsec}) is not compensated by the increase in efficiency (see Tables~\ref{tab:eff1} and \ref{tab:eff2}). Heavy triplets with arbitrary mixing with the SM leptons are excluded for $M \lesssim 90$ GeV in the Majorana case and $M \lesssim 130$ GeV in the Dirac case.

The limits presented in Fig.~\ref{fig:lim1} (Fig.~\ref{fig:lim2}) are completely general  for a single heavy Majorana (Dirac) triplet. However, one should note that in a realistic type-I or type-III seesaw scenario one needs more than one heavy state to generate the light neutrino solar and atmospheric squared mass differences~\cite{Beringer:1900zz}. One possibility is to have two or more heavy triplets. In this, case, the splitting between the two triplet states is expected to be much larger than the intrinsic widths of the heavy leptons (of the order of a MeV for masses $M \sim 200$ GeV and mixings $V_{\ell N}\sim 10^{-2}$), so that interference effects are negligible and the contributions to the signal cross sections, namely the processes in Eqs.~(\ref{ec:pairM}), (\ref{ec:pairD}) for the different triplets, sum. In this case, the limits from experimental searches apply to the lightest triplet, and are conservative. Another possibility is to have one heavy triplet and additional heavy neutrino singlets~\cite{Bajc:2006ia,Dorsner:2006fx,Bajc:2007zf}, for which the general limits on a single heavy triplet apply straightforwardly.

In summary, we have shown that the current CMS search for lepton triplets of type-III seesaw in the trilepton final state~\cite{CMS:2012ra} has good sensitivity even if the mixing with the $\tau$ lepton is dominant. By a reinterpretation of that search, we have shown that general exclusion limits as a function of the triplet mass and mixing can be obtained and, moreover, the results can be compactly presented in two-dimensional plots. We advocate that the results of upcoming experimental searches should be presented in this more general fashion, instead of assumming a fixed arbitrary (and unmotivated) flavour pattern to obtain limits on the triplet mass. Our approach is not only more general than the existing ones, but it also allows to draw the full implications of the experimental searches on the parameter space at a glance. 

\vspace{-2mm}
\acknowledgements
J.A.A.S and F.R.J. thank the CERN Theory division for hospitality during the completion of this work.
This work has been supported by \textit{Funda\c c\~ao para a Ci\^encia e a Tecnologia} (FCT, Portugal) and \textit{Ministerio de Ciencia e Innovaci\'on} (MICINN, Spain) under the bilateral project ``Signals of new fermions at colliders'' (FCT/1683/27/1/2012/S, AIC-D-2011-0811). F.R.J. and P.M.B. acknowledge support from FCT through the projects CERN/FP/123580/2011, PTDC/FIS/102120/2008 and PEst-OE-FIS-UI0777-2013. J.A.A.S. acknowledges support from MICINN through the projects FPA2006-05294 and FPA2010-17915; from Junta de Andaluc\'{\i}a (FQM 101 and FQM 6552) and from FCT (project CERN/ FP/123619/2011).

\end{document}